# Does Outside-In Teaching Improve the Learning of Object-Oriented Programming?

Erica Janke*, Philipp Brune* and Stefan Wagner†
*University of Applied Sciences Neu-Ulm
Neu-Ulm, Germany
Email: {erica.janke, philipp.brune}@hs-neu-ulm.de
†University of Stuttgart
Stuttgart, Germany
Email: stefan.wagner@informatik.uni-stuttgart.de

*Abstract*—Object-oriented programming (OOP) is widely used in the software industry and university introductory courses today. Following the structure of most textbooks, such courses frequently are organised starting with the concepts of imperative and structured programming and only later introducing OOP. An alternative approach is to begin directly with OOP following the Outside-In teaching method as proposed by Meyer. Empirical results for the effects of Outside-In teaching on students and lecturers are sparse, however.

We describe the conceptual design and empirical evaluation of two OOP introductory courses from different universities based on Outside-In teaching. The evaluation results are compared to those from a third course serving as the control group, which was taught OOP the "traditional" way. We evaluate the initial motivation and knowledge of the participants and the learning outcomes. In addition, we analyse results of the end-term exams and qualitatively analyse the results of interviews with the lecturers and tutors.

Regarding the learning outcomes, the results show no significant differences between the Outside-In and the "traditional" teaching method. In general, students found it harder to solve and implement algorithmic problems than to understand object oriented (OO) concepts. Students taught OOP by the Outside-In method, however, were less afraid that they would not pass the exam at the end of term and understood the OO paradigm more quickly.

Therefore, the Outside-In method is no silver bullet for teaching OOP regarding the learning outcomes but has positive effects on motivation and interest.

## I. Introduction

"Nowadays, object oriented programming is the predominant programming paradigm and Java is completely object oriented" [8]. In addition, Java is probably the most used language among IT service providers [15]. Therefore, students frequently learn Java and object oriented programming (OOP) already in introductory courses. This provides a perfect basis for a "later introduction to other paradigms such as logical or functional programming" [11]. Moreover, OOP also helps introducing students to concepts like specification, construction and verification [11].

The classical approach for teaching OOP starts with the basic principles of imperative and structured programming. Teachers introduce variables, data types, control structures, functions or methods and algorithms before the introduction of classes (as an extension of composed data types) and objects. This approach reflects the historical evolution of programming languages and thus the order in which most programmers once have learned these concepts themselves. Yet, this order is meaningless for today's beginners. Moreover, fully object-oriented (OO) languages like Java require the use of OO concepts for all non-trivial (and therefore motivating) tasks. This is especially important for problems which do not have a mathematical character but originate from the field of business information systems.

Therefore, in the last years various approaches for teaching OOP have been proposed following the inverse approach: an "inverted curriculum". These approaches start by introducing objects, interfaces, classes and functions from the beginning [17], e.g. the Outside-In teaching method for OOP based on the programming language Eiffel [12].

*Problem Statement:* Despite several lecturers having already used the Outside-In teaching method in their courses, little empirical evidence on its effects exists. Some authors examined the program compilation behaviour of involved students [16], [9], but its impact on learning outcomes as well as motivation and interest has not been studied so far.

*Research Objective:* Teaching OO concepts to novices is an important task of software engineering education. Therefore, in this paper the feasibility and effects of using Outside-In teaching for teaching OOP in university-level introductory courses using Java are empirically evaluated. In particular, the objective was to understand its impact on learning outcomes as well as on interest and motivation of the participants.

*Context:* We evaluated Outside-In teaching in parallel in two introductory courses for computer science and OOP (Bachelor level, 1st year) at two German universities in the winter term 2012/13. The courses were *Programming and Software Development* (PSD) at the University of Stuttgart (Stuttgart) and *Programming Technique* from the study program "Information Management Automotive" (IMA) at the University of Applied Sciences Neu-Ulm (Ulm). PSD at Stuttgart is part of the curriculum of the Computer Science program as well as of related programs like e.g. software engineering, mechatronics, simulation technology and cybernetics. We used a second course called *Programming Technique* from the program "Information Management and Corporate

Communication" (IMUK) at Ulm as the control group, which was taught in the "traditional" way. The scope and educational objectives of both identically named courses are the same. The lectures at Ulm are given by the same lecturer, who has already taught the IMUK course for several years in an unchanged style. The programming language used in all courses was Java.

*Contribution:* We propose a course design for teaching OOP in introductory courses using the Outside-In method and the Java programming language. We systematically evaluate the application of the proposed course design and teaching method to the two undergraduate courses listed above. The paper presents empirical data resulting from the analysis of two surveys among the students and the evaluation of the exams. We also describe qualitative results from interviews with the involved lecturers and tutors. Our findings indicate that while the approach has its strengths, it cannot be considered as a didactical silver bullet for teaching programming basics as claimed by other authors [12].

## II. RELATED WORK

Originally, the idea of the "inverted curriculum" was proposed by Cohen [5]. It was adopted by various approaches like e.g. the objects-first strategy for introducing OOP. They all follow the "Early Birds" teaching pattern, which requires the introduction of important concepts, the "big ideas", at the very beginning [3]. To support the method, various software tools have been developed which intend to expose students intuitively to using objects (see e.g. [6], [18], [10]).

Meyer picked up the idea of the "inverted curriculum" and developed the Outside-In method for introductory courses of OOP. He used Eiffel as the programming language. In his approach students are provided with high quality software with libraries. Students start as "consumers" of re-usable components and evolve to "producers" of such components [12]. Beside the ETH Zurich, some universities in the USA and Australia successfully apply this concept [6], [18], [10].

A core concept of Outside-In teaching is the use of a sufficiently realistic and complex software system from the very beginning, which students have to get acquainted with and learn to modify and extend [12]. Some authors do not consider the concept of Outside-In teaching but also discuss a similar use of complex software in their lectures. Untch and Offutt [21] describe a system which can be adjusted to different levels of difficulty when teaching software development projects. The application of a software tool implemented in C++ and Eiffel for the introduction to OOP is shown in [14]. However, both systems are not real true-to-life examples for the students. Brügge [4] describes an approach which is more realistic. Their software system enables students to walk through all phases of the software development process, in particular also the phase of system integration. Ramakrishnan et al. [19] present a tool which can mimic huge industrial software projects and thus enables students to understand the benefit of reuse.

Although various lecturers have used similar approaches to Outside-In teaching for some time, only little empirical evidence on its impact can be found in the literature. There are controversial discussions on the effectiveness of the Outside-In method [1] but no empirical results were presented. Cheong Vee, Mannock and Meyer describe a method for evaluating typical errors made by novices in object-oriented programming [16] and Jadud [9] analyses "compilation behaviours of novice programmers". A systematic empirical validation and comparison to the traditional approach is still missing, however.

Therefore, in the following we not only describe a course and software tool design for teaching introductory programming courses Outside-In but also evaluate the approach empirically and compare its outcomes to the traditional course design.

## III. COURSE DESIGN

Based on the approach described in [13], we jointly developed a curriculum for a one-term course. The differences in programming language, scope and the embedding in other lecture courses at the corresponding university were taken into account. The major challenge was the replacement of Eiffel by Java. Eiffel is more puristic in its object-oriented constructs than Java. For example, for Java an early introduction of static methods is necessary because the program is executed with the static main method. The concept of executing a program in Eiffel is considerably easier. Compared to Eiffel, Java lacks a native support for contracts, so additional tools have to be used.

Regarding the scope, the two courses at Ulm and Stuttgart distinctly differ. *Programming and Software Development* (PSD) at Stuttgart consists of four hours lecture and two hours tutorials with teaching assistants per week. *Programming Technique* at Ulm consists of two hours lecture and two hours lecturer-guided tutorials. Furthermore, two additional hours of optional tutorial per week are offered. Some topics like the setup and the functional principle of a computer or data structures are units of separate lecture courses at Ulm. For this reason, these units are not taken into account here. Thus, compared to Ulm, at Stuttgart much more topics are discussed during a semester. PSD at Stuttgart is similar to an introduction to computer science. The aim of the course is a broad introduction to the field of computer science and at least to strike all basic topics. Hence, topics like the principles of computer hardware or the complexity of algorithms are also briefly discussed. Those topics are deepened in corresponding courses like *Algorithms and Data Structures*, *Theoretical Principles of Computer Science* and *Technical Principles of Computer Science*.

The course *Programming Technique* at Ulm is embedded in business-oriented information management study courses and for this reason more oriented towards application and less towards theoretical foundations. Nevertheless, as an important course relevant to information systems in the first term, it has the objective to give at least an overview of the essential basics of information systems development. Another objective is the preparation for courses like *Databases* and *Car IT* as well as

TABLE I
STRUCTURE OF THE COURSES

| | | Ulm | Stuttgart |
|---|---|---|---|
| **Introduction** | After discussing organisational issues, the course starts with a general discussion on computers and programming. We explain the classical "Hello, World" program with the purpose to show students how a program looks like, how it is compiled and executed. With the help of this program, we introduce the Java Virtual Machine. | x | x |
| **Objects** | We introduce object orientation and emphasise thinking in objects. We discuss objects in the real world, abstract objects and pure software objects. We also talk about how they can be used to define the operations of a computer. We introduce classes as blueprints for objects. We also explain primitive data types and typecasts which are necessary to program in Java. | x | x |
| **Execution** | We discuss how objects are created and methods are called. In Java, this means we introduce the *main* method. For that, we present static methods and static attributes. Students learn the structure of a block in the source code and how access can be limited with the help of visibility attributes. There is a short introduction to the integrated development environment (IDE) – in our case Eclipse. In this stage students are able to write first programs by calling a sequence of methods on objects. | x | x |
| **Interfaces** | First, we discuss interfaces of systems and classes. We introduce the concept of contracts to better describe interfaces. As Java has limited possibilities to implement them, they are used as a concept to think about interfaces and to document them. After that, we introduce formal languages using the example of Java and its grammar. | x | x |
| **UML** | We discuss how classes and interfaces can be modelled and documented. For this purpose, we introduce UML class diagrams and UML sequence diagrams. They are used to think about the structuring and interaction of objects. | x | x |
| **Boolean logic** | Boolean logic is essential for computer programming. We define Java's boolean and relational operators as a basis for branching. | x | x |
| **Branches and loops** | Based on the boolean logic, we dive into the implementation of methods. We introduce branches and loops in Java. The set of data types is enlarged by fields, Java Collections and generics. | x | x |
| **Routines** | We introduce the concept of routines and accordingly methods in Java and their meaning. We discuss abstraction, modularisation and information hiding and their representation in Java. Moreover, we talk about exception handling as an additional control structure. | x | x |
| **Variables** | We introduce the concept of variables late. So far, variables have not been mentioned. There were attributes in objects or parameters passed to methods – the core concept are objects. When we implement methods more imperative concepts and, finally, variables are necessary. We then can explain references and memory contents. | x | x |
| **Hardware** | We deepen the knowledge about computers which students gained earlier. What is a computer? Of which parts does a computer consist? Which further hardware is important? | | x |
| **Syntax** | Also building on earlier lectures, we take a deeper look at the grammar of Java described in EBNF and the syntax tree. | | x |
| **Programming languages** | We take a step back from Java. Which other programming paradigms do exist? How do they differ? Which are prominent examples for these paradigms? Also other programming tools like the debugger are explored in more detail. | | x |
| **Data structures and algorithms** | We take a closer look at basic data structures and algorithms and discuss their complexity and efficiency. This does not replace separate lecture courses on algorithms and data structures but lays first foundations. | | x |
| **Recursion** | We introduce recursion as an important concept in programming. As an example data structure, we employ binary trees. | | x |
| **Sorting** | We use simple sorting algorithms as examples for developing a general procedure for the design of algorithms. | x | x |
| **Inheritance and polymorphy** | We complete the object oriented concepts by introducing inheritance and polymorphism. They are introduced late because of their complexity, and they are not necessary for basic object oriented programming. The apply these concepts in Java and the corresponding UML diagrams. We now can also talk about Java interfaces. | x | x |
| **Semantics** | We introduce the concept of semantics of programming languages and their formal description. Students gain insight into formal verification. | | x |
| **GUI** | Students should be enabled to build realistic programs. For this purpose, we introduce graphical user interfaces (Eclipse SWT). The task of the students is to expand the media player. | x | x |
| **Test** | Programming does not only include the initial generation of source code. Another important point is testing by the developer. We cover the development of unit tests and (briefly) test-driven development. | | x |
| **Quality** | This leads to the more general concept of software quality. We discuss the difference between process quality and product quality, quality factors and their influence on programming and the importance of maintenance and evolution. | | x |
| **Software engineering** | The last block represents the direct link to a software engineering course. Students learn which further activities of a development process are necessary to create larger software systems. | x | x |

*Software Engineering* and *Web Engineering* in the second and third semester.

To master such a wide range of topics while being able to support each other, we developed 21 teaching units representing the core topics. One unit consists of one or two double weeks of lectures and the respective tutorials. Table I presents an overview of the different units. The "x" marks whether the unit is covered by lessons at Ulm or Stuttgart. The numbering also reflects the chronological order in the lecture course.

Essential is the introduction to objects. The first five teaching units (the first weeks of the course) solely deal with objects. Only after that we delve into the implementation of methods. Here, branches, loops and finally also variables are introduced. In this phase the classical imperative development is described. This approach should foster the students to think in objects and not in sequences of imperative instructions. After that, we deal with several general topics, depending on the scope of the lecture. Finally, the step from "programming in the small" to "programming in the large" is prepared. In Table I every teaching unit is briefly described.

*Example Software System:* In [13] the Eiffel software package TRAFFIC is used as an example system. This software represents a traffic and tourist information system for visitors in Paris. However, in our setting the usage of this system was not possible because Java is used as the programming language. Because of this, a different example system was developed, which is also closer to the everyday experience and software used by the students than the TRAFFIC system. A MP3-player application was selected, which was realised in Java and is similar to other popular music programs for PCs. The so called "CodeTunes-System" uses the open source library JLayer (see http://www.javazoom.net) for playing back music files and the Standard Widget Toolkit (SWT) (see http://www.eclipse.org/swt) for the implementation of the graphical user interface. Figure 1 shows a screenshot of the main window of the application in its current development state.

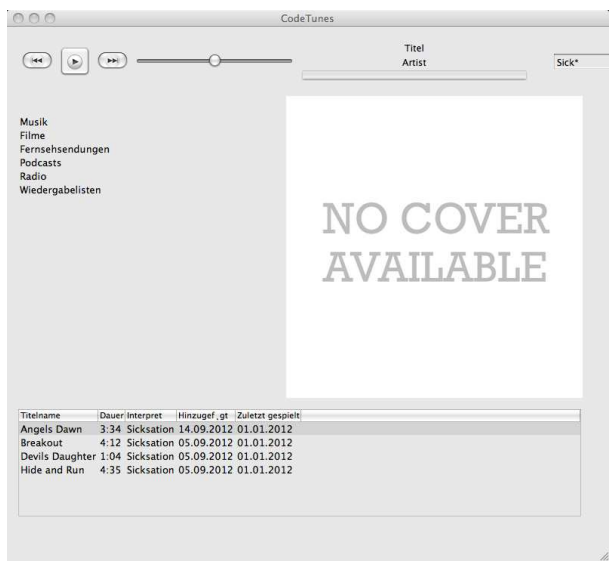

Fig. 1. Screenshot of the Java-based example software system CodeTunes (a media player) developed for the courses.

The selected topic offers enough potential for future projects in further courses, e.g. the implementation and integration of an online music shop. Currently, the software only offers the basic functionality of a media player. During the lessons, students are intended to continually further develop the system, e.g. in the context of student research projects.

Especially at the beginning, in the original TRAFFIC system, the Eiffel classes TOURISM (touristic activity in Paris), STATION (Metro station) and LINE (Metro line) are used as substantial concepts of the application domain [13]. The logical pendants in CodeTunes are the Java classes MediaPlayer (rendering of music titles in various forms), Title (piece of music) and PlayList (a list of pieces of music). Based on this, the programming examples in [13] can be reproduced similarly. For example, in unit 2 students begin with the following simple example class:

```
public class MyFavouriteSongs extends
    SimpleMediaPlayer  {
        public void play()  {
            title1.play();
            title3.play();
        }
}
```

In this class, using title1 and title3, students apply existing objects of the class Title. With the help of those they can play a sequence of music titles.

## IV. CASE STUDY DESIGN

The objectives of our case study are expressed by the following research questions:

RQ 1: How does the self-assessment and motivation of students taught OOP with the Outside-In method differ from those taught with the "traditional" method?

RQ 2: Which learning outcomes does the Outside-In method generate compared to the "traditional" method?

RQ 3: How and to which extent does Outside-In teaching contribute to an improvement of teaching OOP from the lecturers' perspective?

*Data Collection Procedure:* To answer RQ 1, we needed data from the students. Therefore, at the beginning of the term (measuring point I, MPI) we conducted a survey at Ulm and Stuttgart to determine the initial motivation, self-assessment and previous knowledge of the participants. By means of this survey, we were able to take into account the heterogeneous background and personality types of the participants in the two different bachelor programs (Computer Science and related courses at the university in contrast to Information Management at a university of applied sciences). In Germany, universities must select undergraduate students only according to their respective high-school diploma grade and are not allowed to use other selection criteria. Therefore, students are very heterogeneous with respect to their motivation, interest, and study-related skills.

After three weeks of teaching (measuring point II, MPII), we conducted the second survey to measure the achieved learning success and motivation changes of the participants. With the results of this survey, RQ 1 and RQ 2 could be answered. To evaluate the learning outcomes, the second survey contained a test. Students were given a Java class construct and they had to answer different questions about this class. To measure the self-assessment and motivation of the students and their interest in the lectures, we asked the students seven questions which they could answer on a five-point scale. The questions are listed in the catalogue below.

In addition to these surveys, we analysed the results of the end-term exams to extend the empirical basis on learning outcomes and to obtain a broader basis for answering RQ 2.

To obtain a picture of the lecturers' perspective on the efficiency of Outside-In teaching and to answer RQ 3, we interviewed all involved lecturers (two of them are authors) and exercise instructors by means of partially standardised interviews, so the interviewer used a guideline for the interview and the interviewees were free in their answers. Figure 2 shows a quick overview over the different measuring points and the values which were measured at these points.

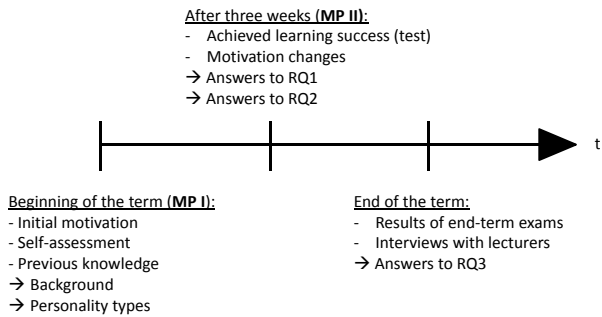

Fig. 2. Timeline overview of the data collection process including measuring points.

The catalogue of questions for the surveys was subdivided into three groups of questions. One for determining motivation and previous knowledge, the second for measuring self-assessment, students' expectations and the perception of the lecture. The third group tested the technical understanding of the principles of OOP. Not all questions were asked at both measuring points. Questions on motivation and previous knowledge were asked on MPI, questions on expectations and perception were asked on MPI and MPII and questions on the learning success were asked on MPII.

The first group of questions for the determination of the knowledge and the motivation included the following questions and possible answers:
1) Have you already completed a job training? (yes / no)
2) Do you have previous knowledge in one or more subjects? (yes / no)
3) I have always been interested in the courses of the study program I have chosen. (applies – does not apply, 4 levels)
4) I think I chose the right study program. (applies – does not apply, 4 levels)
5) I am very interested in the lecture. (applies – does not apply, 5 levels)
6) I will devote myself to the topics of the lecture in my spare time. (applies – does not apply, 5 levels)
7) I will frequently participate in the lecture with questions and comments. (applies – does not apply, 5 levels)

The 4-point scale in questions 3–4 was chosen to force a clear answer. The subjects mentioned in question 2 are not necessarily restricted to IT related topics.

In question group 2 seven questions were asked, to which the participants could respond by choosing one value on a five-point scale from "applies" (= 1) to "does not apply" (= 5):
1) The subject is a challenge for me.
2) I am afraid not to pass the exam.
3) I am devoted to the topics of the lecture in my spare time.
4) I execute my tasks on time.
5) The course has a clear outline.
6) At the end of the semester, my performance will be better than the average.
7) At the end of the semester, I will not have passed.

Question group 3 determine the learning success with respect to basic concepts of OOP in the first weeks of the term. A Java class was presented to the students, who had to answer nine single- or multiple-choice-comprehension questions that referred to different OOP concepts. The tested concepts were "class", "object" and "fundamentals of programming".

The questions in the end-term exams tested different dimensions of knowledge. These dimensions were structured according to [20] in "remember", "understand", "explain", "use", "apply" and "develop". "Remember" means that the student is able to remember information and to reproduce it word for word. The knowledge dimension "understand" implies that the student understands the meaning of information, is able to define it and is able to link new but thematically related information to it. When students are able to identify correlations, dependencies and similarities between information and are able to explain it in their own words, they have reached the dimension "explain". "Use" means that the student is able to apply knowledge in a pre-defined limited context and / or under guidance. Autonomous, independent appliance / application, even in a difficult environment, is described by the knowledge dimension "apply". "Develop" means the creation of new knowledge.

Table II illustrates the structure of the questions in the end-term exam for the course PSD at Stuttgart and Programming Technique at Ulm.

To obtain the lecturers' point of view on the effectiveness of Outside-In teaching, we conducted partially standardised interviews with the two lecturers and two exercise instructors and recorded them.

*Analysis Procedure:* In preparation for the evaluation of the answers to the surveys, we checked the distributions of the answers to the questions of both groups with the help of a Kruskal-Wallis-Test. We chose this test because the measured values are not normally distributed and there are more than two samples. We also checked the results from the exams to determine if the two groups Stuttgart and Ulm IMUK have the same distribution or if the results are significantly different. For this purpose we used the Mann-Whitney-U-Test, because the measured values are not normally distributed but of an ordinal scale and we compared exactly two samples.

For analyzing the answers to questions linked to motivation (i.e. questions 3–7 of question group one), we computed the median of the answers and compared them to the median of the optimal answers. The percentages of the students who answered to questions 1–2 of question group one with "yes" were averaged.

TABLE II
CLUSTERED QUESTIONS OF THE EXAM AT STUTTGART (S) AND ULM (U)
AND THEIR KNOWLEDGE DIMENSIONS.

| Cluster | Issue | remember | understand | explain | use |
|---|---|---|---|---|---|
| Object Orientation | definition of objects | S | S | | |
| | method overloading | S | S | | |
| | interfaces / inheritance | S | S | S | |
| | polymorphism / dynamic binding | S | S | S | |
| | overload / override (inheritance) | S | S | S | |
| | information hiding / static | S | S | S | |
| | class diagrams / Java / algorithms | U | U | | U |
| | class relations | U | U | | U |
| Algorithms | exceptions / exception handling | S | S | S | |
| | static methods | S | S | S | |
| | GUI | U | U | | U |
| | computation of the Fibonacci-numbers | U | U | | U |
| Code Comprehension | static methods / algorithms | S | S | S | |
| | loops | U | U | | U |
| Fundamental Concepts | recursive functions | S | S | S | |
| | loops | U | U | | U |

To evaluate the results of the exam, first we selected the questions related to object orientation. Second, we clustered those questions in four groups, "Concepts of Object Orientation", "Algorithms", "Code Comprehension" and "Fundamental Concepts". We analysed the percentages of points achieved in the exam and compared the results of the involved groups.

The analysis of the qualitative data started with a verbatim transcription of the recorded interviews. Concerning the method, we used selected elements of Grounded Theory following [7]. Where it was not possible to apply elements of Grounded Theory, we substituted them. The number of interviewees was a priori restricted because we examined only two lecture courses. So the sampling followed the a-priori-determination. We chose an inductive approach to coding, which involved open coding, axial coding and selective coding [7].

*Validation Procedure:* In our investigation, we compared two different kinds of universities. Stuttgart represents a German university which aims to provide students with a broad theoretical basis of knowledge and skills. Ulm is a typical German university of applied sciences which focuses more on training students in a practical way for work in industry. This implies different teaching objectives and also different backgrounds of students. To enable a comparison, in our evaluation we considered the motivation and prior knowledge of the students.

The importance of the exams is different at the two universities. In addition to this, students of different majors attend the course at Stuttgart. This fact was given for our experiment and has to be taken into account when interpreting the results. One objective of our investigation was to compare the results for those two different kinds of university the influence of these boundary conditions.

The exams followed different didactic approaches. To enable a comparison, in the presentation of our research results we include the knowledge dimensions tested by the different exercises in the exams.

Two different lecturers taught the lessons, differing in their didactic approach. Nevertheless, the lecturers presented the same basic theoretical knowledge. The technical understanding of OOP was tested in all groups by asking the same questions of question group 3.

To ensure unbiased results in the qualitative research part, two different researchers were involved in the coding of the transcribed interviews. To verify the obtained codes, an additional person was involved to review the codes. The number of interviewees was given a priori and could not be influenced.

For statistical evaluations, the number of participants in the investigated groups has to be comparable. From the Ulm IMA group, only 12 students took part in the exam. At Ulm, students are eligible to postpone exams to the future. Also, at examination time some students already left the study program with the result that not every participating student took part in the exam. So we could not draw any statistical conclusions from that exam.

The background of the students is very different at the two universities. For this reason, we list the most significant differences and deliver statistical data for these cases.

## V. RESULTS

*Course Participants:* The background of the students differs between the two universities. At Ulm, 36.1 % of the IMUK students have already completed a vocational training, at Stuttgart only 6.9 %. At Stuttgart, 95.3 % of the PSD students have a general high school diploma (German "Abitur") for university entrance, while at Ulm only 58.1 % of the IMUK students and 57.6 % of the IMA students have such a graduation. There, 33.3 % (IMUK) and 42.4 % (IMA) of the students qualified for their studies with a university of applied sciences entrance qualification, which is a more technical certificate. 33.3 % of the PSD students at Stuttgart state to aim at a master's degree. Only 18.3 % of the Ulm IMUK students and 3.2 % of the Ulm IMA students state this.

Also the gender distribution is not equal in the three study courses. At Stuttgart, 18.5 % of the PSD students are female. At Ulm, in the study course IMA 27.3 % of the Programming Technique students are female, in the study course IMUK this number amounts to 71.7 %. These results are summarized in Table III.

TABLE III
BACKGROUND INFORMATION OF THE STUDENTS CLUSTERED BY THE THREE DIFFERENT STUDY COURSES STUTTGART PSD (STUTTGART), ULM IMA (HA) AND ULM IMUK (HU, CONTROL GROUP).

| Background Information | Stuttgart | HA | HU |
|---|---|---|---|
| Completed vocational training | 6.9% | 21.2% | 36.1% |
| General high school diploma | 95.3% | 57.6% | 58.1% |
| University of applied sciences entrance qualification | 0.3% | 42.4% | 33.3% |
| Aim at master's degree | 33.3% | 3.2% | 18.3% |
| Female students | 18.5% | 27.3% | 71.7% |

*Qualitative Experiences of the Lecturers (RQ3):* The analysis of the interviews revealed interesting results. Depending on the role – lecturer or exercise instructor – different impressions were gained in some parts of effectiveness of the Outside-In method.

The main categories, which arose from coding, were the following ones (also shown in Figure 3):

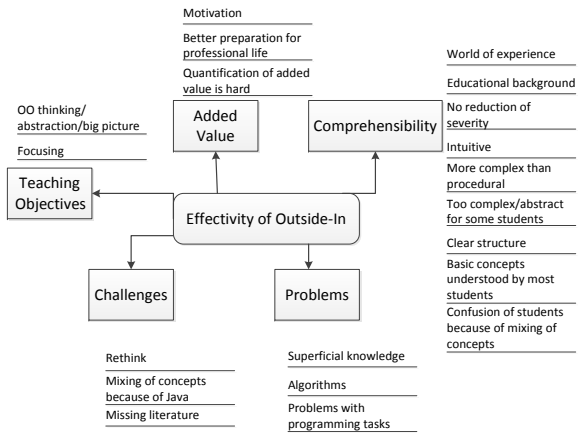

Fig. 3. Categories resulting from open, axial and selective coding and the corresponding categories of the first tier.

*Teaching objectives:* All four experts agreed that with the help of Outside-In teaching students gain a better understanding of complex systems and are more able to deal with such systems. They receive a realistic picture of software engineering. Students are forced to think on a more abstract level. The Outside-In method also enables a better and sustainable understanding of the object oriented perspective. Students gain a better understanding and feeling for objects. Another result is described by the following statement: "I approve to start with this way of thinking. Not starting with the procedural way, which might come naturally to them. But in a long-term view, the OO way is the lasting approach".

*Added value:* As added value from using the Outside-In method, the interviewees listed motivation and a better preparation for professional life. "Objects First enables impressive results of the own programming after a relative short time. ... I think, this contributes to a higher motivation." It was also mentioned that it is hard to quantify an added value of the usage of the Outside-In teaching method.

*Comprehensibility:* In the lecturers' point of view this teaching method is especially appropriate for students who have not yet gained any experience with programming. The exercise instructors had the opposite impression - the Outside-In method makes it harder for students without any prior programming experience to immerse themselves in the object oriented paradigm. The interviewees also mentioned the applicability for all students because of the connection to their world of experience. The lecturers as well as one exercise instructor got the impression that with the help of the Outside-In method, students gained a better understanding of the object oriented paradigm in a comparatively short time. The concept is intuitive, for lecturers as well as for students. It delivers a clear structure, in which structural concepts have to be addressed. Some interviewees had the impression that for some students this structure is too complex and deterrent. They also had the impression that students are confused because of the mixing of concepts during explaining. Another point mentioned was: "I do not think it makes it easier for students. It becomes different, maybe more meaningful, but it does not become easier."

*Challenges:* The lecturers themselves were forced to rethink basics of the object-oriented paradigm because they had to explain it in a different, unfamiliar way. Both lecturers mentioned challenges in explaining concepts. The Outside-In method requires a fixed order of topics. In combination with Java, this implies that certain concepts have to be explained using concepts that are going to be explained in detail later. One lecturer had the impression that this led to confusion for some students. Another challenge is the lack of textbooks which treat the object oriented paradigm in combination with Java and the Outside-In method.

*Problems:* Not enough time is spent on basic programming. Problems occur when students have to develop their own algorithms and e.g. think about how to efficiently use the storage. Students are overwhelmed with the task of building an algorithm. "I think, next time I will invest more time in strengthening algorithmic thinking." And students are over-challenged when they have to think of basic programming concepts.

*Quantitative Results of the Surveys:* At Stuttgart, 237 students attended at MPI and 253 at MPII. They were aged between 17 and 30 years. The lectures at Ulm have been visited by 154 students at MPI and by 117 at MPII, which were aged between 18 and 25 years. Of these, there were 33 students at MPI and 19 at MPII in the study course IMA and 121 at MPI as well as 98 at MPII in the study course IMUK (control group). To illustrate the prior knowledge of the participants, the arithmetic average of the answers to question 1 and 2 from the first group of questions has been used. The motivation was represented by the median of questions 3 to 7 from the first group of questions. For all three groups (Stuttgart PSD, Ulm IMUK and Ulm IMA) the median to those questions is 2. Calculating the median of the optimal answers (completely agreement) would result in a median of 5. The calculated indicators for prior knowledge at MPII are plotted in Figure 4 for IMA, IMUK and Stuttgart PSD.

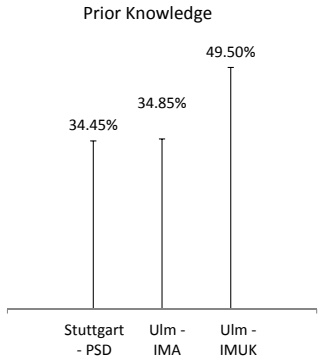

Fig. 4. Proportion of students with previous knowledge at MPI in the three courses.

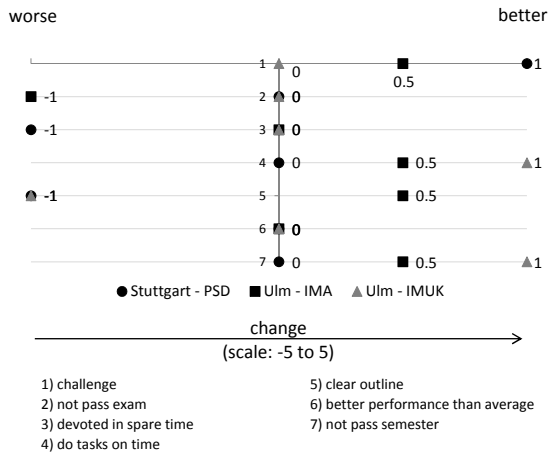

Fig. 5. Changes of self-assessment and the course perception in the three groups of participants between MPI and MPII.

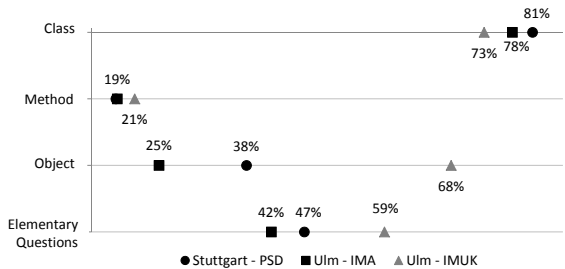

Fig. 6. Percentage of correct answers to the questions about basic concepts of OOP (question group 3) at MPII.

Self-assessment and perception of the course were measured at MPI and MPII. The differences between the medians at MPI in relation to MPII are shown in Figure 5 (comparison of IMA, IMUK and the University of Stuttgart). The understanding of basic concepts of OOP was only measured at MPII. The answers are a snap-shot of the learning success of the students and allow a direct comparison of Outside-In with the "traditional" approach. Figure 6 shows the number of right answers to the questions related to different OOP concepts for each course.

TABLE IV
RESULTS OF THE KRUSKAL-WALLIS-TEST FOR INDEPENDENT SAMPLES TO VERIFY IF THE DISTRIBUTIONS OF THE ANSWERS TO THE QUESTIONS LINKED TO MOTIVATION AND PRIOR KNOWLEDGE ARE THE SAME FOR STUTTGART PSD, ULM IMUK AND ULM IMA WITH CONFIDENCE INTERVAL 95 % AND SIGNIFICANCE LEVEL 0.05.

| category | question | significance | accept / reject |
|---|---|---|---|
| motivation | interest in study course | 0.000 | reject |
| | interest in lecture | 0.000 | reject |
| | study course was a good decision | 0.795 | accept |
| | self study in spare time | 0.137 | accept |
| | participation in lecture | 0.000 | reject |
| prior knowledge | completed vocational training | 0.000 | reject |
| | knowledge in one or more subjects | 0.290 | accept |

Statistical tests reveal the distribution of the answers to questions related to motivation and prior knowledge. The null hypothesis tested was "The distribution of the answers to a single questions is similar for the different study groups Stuttgart PSD, Ulm IMUK and Ulm IMA." For testing this hypothesis, we used the Kruskal-Wallis-test for independent samples. The chosen confidence interval was 95 %. Table IV shows the results.

*Quantitative Results of the Exams:* At Stuttgart, 449 students took part in the exam. At Ulm, in the study course IMUK – which is the control group – 144 students took part in the exam. In the study course IMA, only 12 students took part in the exam. This number is too small to make reliable statistical statements. Because of this, the results of the exam of Stuttgart and the control group at Ulm are compared. Figure 7 shows the question clusters and corresponding percentage of all achieved points with respect to the achievable points.

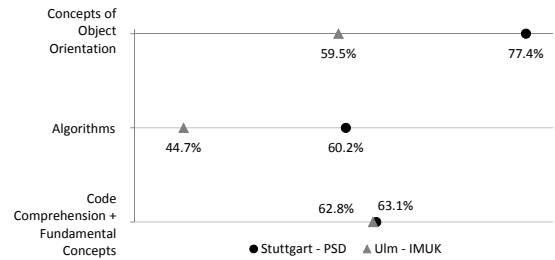

Fig. 7. Percentage of achieved points in the exams for the two courses Stuttgart PSD and Ulm IMUK sorted by question cluster and size.

*Analysis of results:* (RQ1) The survey results presented in Figure 4 show that in both courses (PSD at Stuttgart and IMA at Ulm) the proportions of students with previous knowledge are of similar size. However, this proportion of students is significantly higher in the control group of the IMUK students. The subjects students mentioned to have previous knowledge in are not necessarily IT-related. The motivation of all test groups is on a comparable and relatively high level. IMA students tend to perceive the course as rather demanding. This feeling increases during the first weeks of the lectures as the change in the responses to question 2 indicates (see Figure 5). The rating of the lecture difficulty developed positively in both groups – the PSD students at Stuttgart as well as the IMA students at Ulm – compared to the control group, as the change in the answers to question 1 in Figure 5 indicates. Moreover, students who are taught using the Outside-In method seem to perceive the lecture as more clearly structured than the control group, which is shown by the change in the answers to question 5 in Figure 5. The IMA students did not change their perception at all, whereas the perception of the control group significantly became more negative. In accordance with the lecturers, the students perceived the lecture as better structured and more logical when being taught using the Outside-In method. The differences between Ulm and Stuttgart could be reasoned in the two different instructors.

(RQ2) Considering the learning outcomes, the results for the third question group show that the proportion of correct answers does not differ much between all three groups. In the first question group, only the control group differed strongly with respect to one question in the first category. Also a strong difference can be observed for one question in the third category. One reason for the learning achievements of the IMUK students could be the higher proportion of students with previous knowledge.

The survey also offered the possibility of a direct comparison between the learning outcomes of students at a University and a University of Applied Sciences. Interestingly, no characteristic differences were found.

The results of the exams show that students taught OO using the Outside-In method predominantly understand the OO concepts and are able to explain parts of it. In seven questions related to OO concepts students achieved 77.4 % of the achievable points. In five out of seven questions they achieved more than 75 % of the points. Only run-time exceptions and exception handling seemed to be more difficult, here 68.3 % of the possible points were achieved. Regarding polymorphism and dynamic binding students achieved 57.2 % of all possible points, which still is more than the half. Considering the "Algorithms" part, students seem to have some difficulties in creating algorithms, they achieved 60.2 % of the achievable points. This impression is supported by the results for the "Code Comprehension" part. Here students gained 58.2 % of the points. Understanding a given algorithmic program and creating and implementing new algorithms seem to be the main challenges. The fundamental concepts part seemed to be easier for the students. Here they achieved 67.0 % of the maximum points.

The same is observed in the control group. Whenever students had to create and program algorithms on their own, they achieved only 44.7 % of the maximum points. The amount of achieved points with respect to OO concepts is 59.5%, which indicates the understanding of the OO paradigm is also still improvable. However, when we compare the results of the two groups, we have to keep in mind the different knowledge dimensions tested in the exams at Stuttgart and Ulm.

The interviews revealed that by the Outside-In teaching method students gain a quicker understanding of the OO paradigm. Their perspective is OO from the very beginning, they are able to think in components after a very short time. At the same time, they have problems when asked to build an algorithm and write a program for it, i.e. when it comes down to the very basic concepts of programming.

In general, the observed effects are relatively small and require further empirical examination. With this first investigation there is currently no statistical significance. However, we observe a tendency in the qualitative impressions of the lecturers that has to be verified by further studies over a longer time interval.

*Evaluation of Validity:* One significant threat to validity was the difference of the two compared universities and the students involved in the surveys. To enable a comparison of the results, we presented the percentage of students with prior knowledge and their motivation. We also described the course designs and the different types of universities which were compared and their roles in the educational landscape. We delivered statistical data on completed vocational training, qualification for studies, intended degree and gender distribution to support the analysis of the results.

To be able to compare the results of the exams having different didactic approaches, we clustered the questions and identified the knowledge dimensions tested by the different exercises.

We evaluated the distributions of both, the answers to the questions on motivation and prior knowledge as well as the results of the exams. This revealed that the three groups are very heterogeneous to a large extent. The only observables which had the same distribution for all three courses were prior knowledge, the choice of the study program and the amount of self-studying in spare time.

For the qualitative analysis of the interviews, two researchers coded the transcribed interviews to keep bias to a minimum.

## VI. CONCLUSION

Overall, the use of Outside-In teaching can be seen as successful.

For the teacher, it sometimes is unusual and requires discipline in the implementation and sometimes argumentative tricks need to be used not to anticipate some concepts. But in general, it does not seem to be an impossible task. In return, we have seen many positive aspects, such as a more rigorous representation and working with realistic programs.

Students as well have predominantly seen the structure of the courses as understandable. The self-assessment of students, taught OO with the Outside-In method did not significantly differ from the self-assessment of students who were taught OO the "traditional" way.

Results from the test in the second survey and results of the exam did not show any significant differences of the two groups. The lecturers had the opinion, that students who were taught OO with the Outside-In method had a quick understanding and feeling of the object-oriented paradigm.

A clear statement of whether it benefits the specific learning, cannot be delivered, yet.

*Relation to Existing Evidence:* In the actual state, our results confirm the statement of Schiaffino: "I have found that, for the most part, the greatest obstacle for students in CS1 is learning to analyze a problem and design an algorithm for its solution. ... Whether you use objects first or not does not seem to be particularly important." [2] We also saw that students had problems in building algorithms. The understanding of OO did not significantly differ between the groups taught with and without Outside-In. Further investigations will need to show, if this result holds.

*Impact/Implications:* Our results show, that the introduction of OO with the help of Outside-In teaching does not reduce the teaching results. Students seem to have a quick understanding of the object oriented paradigm. To obtain a clearer picture of the implications, we need replications and extensions of this study.

*Limitations:* Our investigations encompassed only three courses at two universities. One course had a small number of students. Hence, the generalisation is difficult. The two universities are of a different kind and so are the students enrolled in these universities. They have a different qualification, different background. The comparison of the results must respect this. Also two different lecturers with their personal didactic approach taught the students.

*Future Work:* In the following semesters, we will gather more data to be able to evaluate the Outside-In teaching method on a broader base. We will also collect the impressions of the students. The future will then provide a wider base of data.

ACKNOWLEDGMENTS

The present work as part of the EVELIN project was funded by the German Federal Ministry of Education and Research under grant number 01PL12022E. The authors are responsible for the content of this publication.

## REFERENCES

[1] F. Bailie, G. Blank, K. Murray, and R. Rajaravivarma. Java visualization using BlueJ. *Journal of Computing Sciences in Colleges*, 18(3):175–176, 2003.

[2] F. Bailie, M. Courtney, K. Murray, R. Schiaffino, and S. Tuohy. Objects First - does it work? *Journal of Computing Sciences in Colleges*, 19(2):303–305, 2003.

[3] J. Bergin. Fourteen pedagogical patterns, 12.01.2006.

[4] B. Bruegge. Teaching an industry-oriented Software Engineering course. In *Proceedings of the Software Engineering Education Conference, SEI 1992*, volume 640 of *Lecture Notes in Computer Science*, pages 63–87. Springer, 1992.

[5] B. Cohen. The education of the Information Systems engineer: To meet the present and future needs for staffing of the IT industry with appropriately qualified and trained people, new and radical approached to education and training are required. *Electronics and Power*, 33(3):203–205, 1987.

[6] S. Cooper, W. Dann, and R. Pausch. Teaching Objects-First in introductory Computer Science. In *Proceedings of the 34th SIGCSE Technical Symposium on Computer Science Education*, volume 35, pages 191–195, 2003.

[7] J. M. Corbin and A. L. Strauss. *Basics of Qualitative Research: Techniques and Procedures for Developing Grounded Theory*. Sage Publications, 3 edition, 2008.

[8] C. S. Horstmann and G. Cornell. *Core Java 2 Vol.1: Fundamentals*. Prentice Hall, 8th edition, 2007.

[9] M. C. Jadud. A first look at novice compilation behaviour using BlueJ. *Computer Science Education*, 15(1):25–40, 2005.

[10] M. Kölling and J. Rosenberg. Guidelines for teaching Object Orientation with Java. In *Proceedings of the 6th Annual Conference on Innovation and Technology in Computer Science Education*, volume 33, pages 33–36, 2001.

[11] B. Meyer. Towards an Object-Oriented curriculum. In *Proceedings of 11th International TOOLS Conference*, pages 585–594, 1993.

[12] B. Meyer. The outside-in method of teaching introductory programming. In *Perspectives of Systems Informatics, 5th International Andrei Ershov Memorial Conference, PSI 2003*, volume 2890 of *Lecture Notes in Computer Science*, pages 66–78. Springer, 2003.

[13] B. Meyer. *Touch of Class: Learning to Program Well with Object Technology, Design by Contract, and Steps to Software Engineering*. Springer, 2003.

[14] C. Mingins, J. Miller, M. Dick, and M. Postema. How we teach Software Engineering. *Journal of Object-Oriented Programming*, 11(9):64–66, 74, 1999.

[15] A. Neumann. Umfrage: Java ist verbreitetste Programmiersprache bei IT-Dienstleistern (in German), 2011.

[16] M. H. Ng Cheong Vee, B. Meyer, and K. L. Mannock. Empirical study of novice errors and error paths in Object-Oriented programming. In *Proceedings of the 7th Annual HEA-ICS Conference*, 2006.

[17] M. Pedroni and B. Meyer. The Inverted Curriculum in practice. In *Proceedings of the 37th SIGCSE Technical Symposium on Computer Science Education*, volume 38, pages 481–485, 2006.

[18] V. K. Proulx, J. Raab, and R. Rasala. Objects from the beginning — with GUIs. In *Proceedings of the 7th Annual Conference on Innovation and Technology in Computer Science Education*, volume 34, pages 65–69, 2002.

[19] S. Ramakrishnan, C. Mingins, B. Henderson-Sellers, and R. Duke. Planned software reuse in Object-Oriented Software Engineering education. In *Proceedings of the 1994 Software Education Conference*, pages 250–254, 1994.

[20] Y. Sedelmaier, S. Claren, and D. Landes. Welche Kompetenzen benötigt ein Software Ingenieur? (in German). In H. Lichter and A. Spillner, editors, *Proceedings of the SEUH 2013*, pages 117–128, 2013.

[21] R. H. Untch and A. J. Offutt. Integrating research, reuse, and integration into Software Engineering course. In *Proceedings of the Software Engineering Education Conference, SEI 1992*, volume 640 of *Lecture Notes in Computer Science*, pages 88–98. Springer, 1992.